\documentclass[conference]{IEEEtran}
\IEEEoverridecommandlockouts
\usepackage{cite}
\usepackage{amsmath,amssymb,amsfonts}
\usepackage{algorithmic}
\usepackage{graphicx}
\usepackage{textcomp}
\usepackage{xcolor}

\usepackage[binary-units]{siunitx}
\DeclareSIUnit{\dbm}{dBm}

\begin{document}

\title{On the Dynamic Range of Digital Correlative Time~Domain Radio Channel Measurements}

\author{\IEEEauthorblockN{Sven Wittig, Wilhelm Keusgen and Michael Peter}
\IEEEauthorblockA{\textit{Fraunhofer Heinrich Hertz Institute} \\
Berlin, Germany \\
\{sven.wittig, wilhelm.keusgen, michael.peter\}@hhi.fraunhofer.de}
\and
\IEEEauthorblockN{Taro Eichler}
\IEEEauthorblockA{\textit{Rohde \& Schwarz} \\
Munich, Germany \\
taro.eichler@rohde-schwarz.com}
}

\maketitle

\begin{abstract}
For the concise characterization of radio channel measurement setups, different performance metrics like dynamic range and maximum measurable path loss have been proposed. This work further elaborates on these metrics and discusses practical bounds for the class of digital correlative time domain channel sounders. From a general instrumentation model, relevant nonidealities and the resulting effects on measurement performance are identified. These considerations yield the tools to select appropriate measurement parameters for a given scenario and assess the overall performance of a particular measurement setup. The findings are further illustrated with data acquired from an instrument-based measurement setup.
\end{abstract}



\section{Introduction}
In the advancement of wireless communication technology, making new portions of the radio spectrum available for technological exploitation is a key approach in the endeavour of meeting increased performance demands. Recently, the research towards fifth generation (5G) mobile radio networks and beyond has been targeting frequencies ranging from \SI{6}{\giga\hertz} up to \SI{300}{\giga\hertz}. The first step on the path to conquering new frequency bands is a thorough characterization of the radio propagation channel at the frequencies of interest by measurement, ultimately leading to the development of channel models that facilitate system- and link-level simulations. Consequently, channel sounding measurements, especially in the millimeter-wave and sub-THz frequency ranges, have received high interest during the last years.

With a large number of different channel measurement techniques and practical realizations thereof in existence \cite{salous2013, maccartney2017}, ensuring validity and comparability of the acquired measurement data remains a challenge. However, this is a necessary condition for the creation of a statistically robust basis of measurement data for the derivation of channel model parameters.

Focusing on the path loss measurement capabilities of channel sounders, the contribution of this work is twofold: Firstly, we give clear definitions of general performance metrics that are independent of a particular technical realization. Secondly, for the class of digital correlative time domain channel sounders, we determine practical bounds on these metrics by simulation using a general instrumentation model. These results are then verified by practical measurements on an instrument based channel sounder setup.


\section{Performance Metrics}
Typical performance metrics for the characterization of channel sounders include accuracy, precision, resolution and stability with respect to each of the measurement quantities, e.g. delay, amplitude or angle of arrival \cite{peter2016}. Additionally, the technical realization of a particular sounding technique imposes limits on the range of values of these quantities that may be captured by it. All these metrics are mostly determined by nonidealities such as -- among others -- thermal noise, phase noise or nonlinear effects in the signal chain.

In the following discussion, we will focus on the performance of channel sounders with respect to the measurable range of amplitudes, i.e. path losses, in an individual captured channel impulse response (CIR). To that end and elaborating on \cite{peter2016}, three metrics are defined to concisely describe the amplitude measurement capabilities.

\subsection{Dynamic Range}
\label{sec:dynamic-range}
The dynamic range of a CIR measurement is defined as the ratio of the strongest multipath magnitude to the 0.99-quantile ($Q_{0.99}$) of error contribution magnitudes, usually expressed in \si{\decibel}. Choosing $Q_{0.99}$ as a measure for the CIR noise floor instead of, for example, the maximum value gives the dynamic range metric a meaningful statistical interpretation. Following this definition, the dynamic range itself is the 0.01-quantile of peak-to-error magnitude ratios in a CIR.

To clearly differentiate between the discussion of individual measurements vs. overall performance of the measurement setup, the dynamic range determined from a particular CIR is denoted the \textit{Instantaneous Dynamic Range} $\mathit{IDR}$ of that CIR. When discussing dynamic range as a figure of merit for channel sounders it is denoted the \textit{Achievable Dynamic Range} $\mathit{DR}_A$. $\mathit{DR}_A$ depends on the particular parametrization of the measurement setup and it holds $\mathit{IDR} \leq \mathit{DR}_A$. The maximum achievable dynamic range among all possible parametrizations of a channel sounder is denoted $\mathit{DR}_{A,0}$.

\subsection{Minimum Measurable Path Loss}
The minimum measurable path loss $\mathit{PL}_\mathrm{min}$ is defined as the inverse of the strongest magnitude in a single-tap CIR that may be measured without reaching receiver saturation, expressed in \si{\decibel}. With this definition, $\mathit{PL}_\mathrm{min}$ is a deterministic quantity depending on transmitter and receiver linearity of the measurement system.

It depends on the parametrization of the measurement setup and may be arbitrarily set by an appropriate choice of RF gain (i.e. transmit power) up to an upper bound $\mathit{PL}_{\mathrm{min},0}$ which is reached for transmitter and receiver operating at the edge of their linear region.


\subsection{Maximum Measurable Path Loss}
The maximum measurable path loss $\mathit{PL}_\mathrm{max}$ is defined as the sum of minimum measurable path loss and achievable dynamic range for a particular parametrization of the measurement setup,
\begin{equation}
\mathit{PL}_\mathrm{max}=\mathit{PL}_\mathrm{min}+\mathit{DR}_A. 
\end{equation}
As a channel sounder figure of merit, the maximum achievable maximum measurable path loss among all parametrizations is given by
\begin{equation*}
\mathit{PL}_{\mathrm{max},0}=\mathit{PL}_{\mathrm{min},0}+\mathit{DR}_{A,0}.
\end{equation*}


\section{Measurement Principle}
Channel sounding techniques may be broadly divided into frequency domain and time domain approaches, that make different trade-offs with respect to parameters such as measurement duration and dynamic range \cite{salous2013}. In time domain approaches, the channel is excited with a wideband sounding signal to record the CIR at the receiver. To circumvent the problems associated with direct pulse excitation, most channel sounders make use of low crest factor pulse compression waveforms that yield the CIR by means of a correlation operation at the receiver.

Pulse compression offers the benefit of additional processing gain which is determined by the product of signal bandwidth $B$ and measurement time $T_m$, usually expressed in \si{\decibel}:
\begin{equation}
    G_\mathrm{proc} = 10\cdot\log_{10}(BT_m)
\end{equation}
It is noted that this is an upper bound on processing gain which may not be fully realized by a particular technical realization. For periodic pulse compression sounding signals, $G_\mathrm{proc}$ may be further decomposed into correlation gain depending on signal period $T_{p}$ and the gain from averaging $K$ periods of the received signal:
\begin{equation}
\begin{split}
    G_\mathrm{proc} &= G_\mathrm{corr}+G_\mathrm{avg} \\
           &= 10\cdot\log_{10}(BT_{p})+10\cdot\log_{10}(K)
\end{split}
\end{equation}

In digital correlative channel sounders, the receiver performs wideband sampling of the received signal with the correlation implemented in the digital domain, either in real-time or during post processing. This facilitates the use of complex-valued sounding signals. Complex-valued signals offer the possibility to create sequences with maximum power efficiency and perfect correlation properties. Choosing a set of random complex variables with uniform  magnitude and applying discrete inverse Fourier transform allows for an infinite number of perfect correlation sequences of given length, whereas for these sequences the peak-to-average power ratio (i.e. crest factor) could not be controlled. Therefore, specific sequences with dedicated numbers of signal states (phase values) and crest factors bounded to low values are favourable.

In 1972 Chu introduced complex poly-phase sequences of arbitrary length $N$ and parameter $\lambda$, which are uniform and have perfect periodic correlation properties \cite{chu72}. These sequences were later known as Frank-Zadoff-Chu sequences (FZC sequences) and were widely adopted in communication engineering. One sequence $\{s^{N}_{\lambda,n}\} = \{s^{N}_{\lambda,0}, s^{N}_{\lambda,1}, \dots ,s^{N}_{\lambda,N-1}\} $ is defined by $N$ and $\lambda$ whereas both have to be relatively prime:
\begin{align*}
1 \leq \lambda &\leq N-1  &&  \{N , \lambda\}  \in \mathbb{N}\\
 \mathrm{gcd} (N,\lambda) &= 1&&
\end{align*}
The phases $\xi_n$ of a sequence $\{s^{N}_{\lambda,n}\}$ are given by
\begin{align}
\xi^{N}_{\lambda,n} &=\pi \frac{\lambda}{N} n  (n + 1) && n = 0 \dots N-1 && \text{for odd $N$}\\
\xi^{N}_{\lambda,n} &=\pi \frac{\lambda}{N}  n^2  && n = 0 \dots N-1 &&  \text{for even $N$}
\end{align}
The sequence $\{s^{N}_{\lambda,n}\}$ is calculated as the complex exponential of $\{\xi^{N}_{\lambda,n}\}$ (with $j = \sqrt{-1}$ being the imaginary unit):
\begin{equation}
s^{N}_{\lambda,n} = \exp(j \xi^{N}_{\lambda,n})
\end{equation}

Due to their perfect periodic correlation properties, FZC sequences are often chosen as the pulse compression waveform for digital correlative channel sounder implementations. However, it must be noted that the perfect properties are lost under the impairments encountered in an analog signal chain. The impaired cross correlation function exhibits non-negligible sidelobes, hence limiting the dynamic range of the measurement system.

An important property of the cross correlation of impaired FZC sequences with respect to the definition of dynamic range given in Sec.~\ref{sec:dynamic-range} is its \textit{peak-to-sidelobe ratio (PSR)}. For our purposes, PSR is defined as the ratio of cross correlation peak magnitude to the 0.99-quantile of all other cross correlation function sample magnitudes, expressed in \si{\decibel}. With this definition, PSR corresponds directly to the dynamic range of a single-tap CIR measured by correlation.

\begin{figure}[htbp]
\centerline{\includegraphics[width=3.375in]{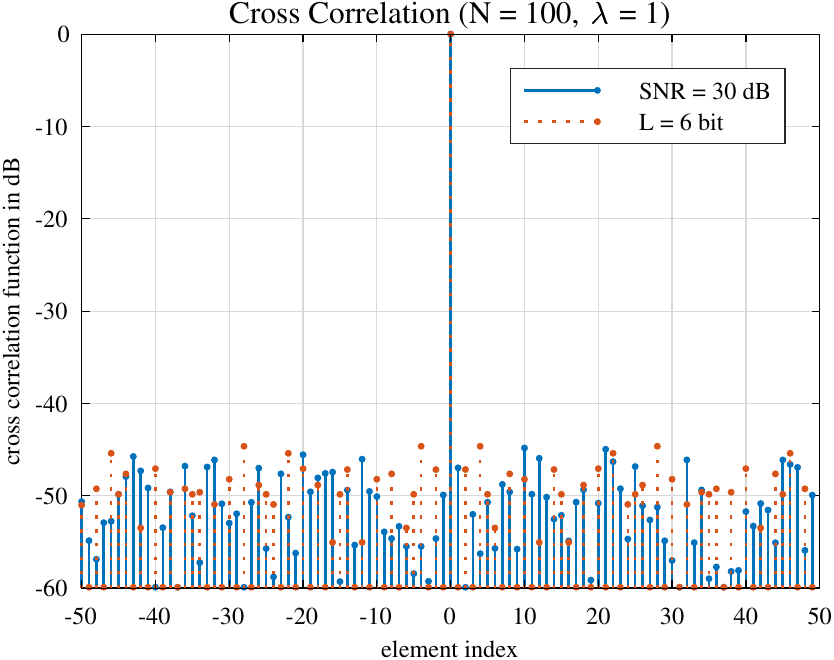}}
\caption{Normalized cross correlation of FZC sequence $s^{100}_{1}$ with $s^{100}_{1}$ impaired by \SI{-30}{\decibel} AWGN or \SI{6}{\bit} quantization, respectively. In this plot, values below \SI{-60}{\decibel} are set to \SI{-60}{\decibel}.}
\label{fig:akf_impaired}
\end{figure}

To illustrate the degradation in sidelobe level for typical impairments, Fig.~\ref{fig:akf_impaired} shows the periodic cross correlation of a sequence $s^{100}_{1}$ and a copy of it impaired by either \SI{-30}{\decibel} additive white gaussian noise (AWGN) or by \SI{6}{\bit} quantization. Being a deterministic effect, quantization results in a symmetric sidelobe structure with bounded $\mathit{PSR}$, while additive noise results in a non-deterministic and unbounded noise floor best described by a statistical $\mathit{PSR}$ as defined above.

The properties of FZC sequences transmitted over analog signal chains, including PSR degradation, generally depend on the root parameter $\lambda$ and sequence length $N$. Hence, we will limit our following discussion to the case of $\lambda=1$ without loss of generality.


\section{Instrumentation Model}
\begin{figure}[htbp]
\centerline{\includegraphics[width=3.375in]{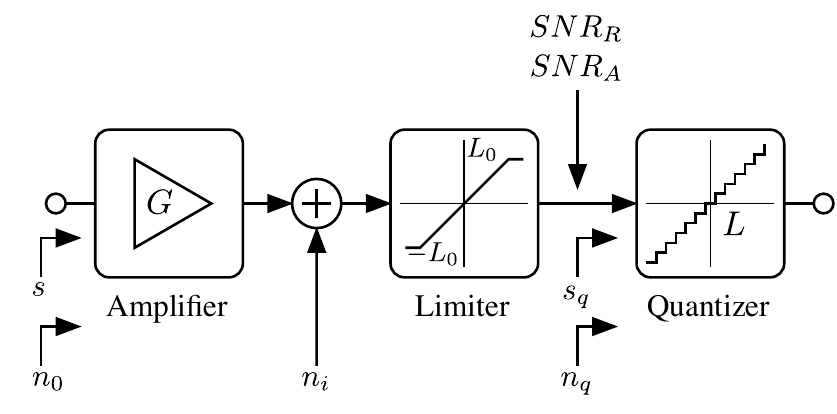}}
\caption{Generalized digital receiver model.}
\label{fig:frontend}
\end{figure}

For systematic investigation of the $\mathit{PSR}$ degradation resulting from nonidealities in the analog signal chain, the generalized digital receiver model of Fig.~\ref{fig:frontend} is considered.

The receiver is presented with an input signal $s$ with bandwidth $B$ and signal power $S$. The relevant noise contributions and nonlinear effects affecting the receiver output are captured by this model as follows:
\begin{itemize}
    \item \textit{Thermal noise} $n_0$: Thermal noise is referred to the receiver input with noise power $N_0=\SI{-174}{\dbm\per\hertz}+10\cdot\log_{10}(B)$
    \item \textit{Intrinsic receiver noise} $n_i$: All noise contributions by the receiver are lumped into the intrinsic receiver noise signal with power $N_i$, often charaterized by the receiver noise figure $\mathit{NF}$.
    \item \textit{Phase noise}: While phase noise is an important factor for the overall system performance, it is omitted from the model in this work and left for an isolated discussion. However, this does not affect validity of the results for dynamic range in the sense of an upper bound, i.e. the main effect of phase noise w.r.t. the following discussion is expected to be a degradation of $G_\mathrm{proc}$. A detailed investigation of FZC cross correlation properties under phase noise is subject to future work.
    \item \textit{Amplitude limiting}: The finite linear region of the receiver is modelled by an amplitude limiting operation, resulting in a maximum input signal level $S_\mathrm{max}$. It follows that for signals with crest factors other than \SI{0}{\decibel}, sufficient backoff must be considered. The analog crest factor of FZC sequences with root parameter $\lambda = 1$ and $\lambda = N-1$ with certain, relevant lengths ($N > 100$) is bounded by \SI{2.6}{\decibel}, hence the intrinsic backoff of signal chains dimensioned using sinusoid signals (crest factor \SI{3}{\decibel}) as a reference is sufficient. Since the maximum magnitude of real or imaginary part cannot exceed the magnitude of a complex signal, this also holds in the practical case of IQ-sampling. No additional backoff must be included to accommodate the additive noise, since $GS \gg GN_0+N_i$ holds at the edge of the limiter range. From a theoretical perspective, the crest factor of the received signal is increased by the channel frequency response during a real channel sounding situation, but in most practical cases this does not need to be considered, since the full scale level at the receiver is typically reached for a strong line-of-sight situation with a relatively smooth frequency transfer function.
    \item \textit{Quantization}: The received signal after amplification, noise and limiting $s_q$ is quantized with \SI[number-math-rm=\mathnormal, parse-numbers=false]{L}{\bit} resolution, resulting in an LSB of $\Delta=2^{1-L}L_0$.
\end{itemize}

When operating within the receiver's linear region, all nonidealities prior to quantization are quantified by the \textit{received SNR}, $\mathit{SNR}_R=\tfrac{S_q}{N_q}$. Its upper bound is given by the \textit{achievable SNR},  $\mathit{SNR}_A=\tfrac{S_{q,\mathrm{max}}}{N_q}$, which is reached for the maximum linear input signal level $S_\mathrm{max}$, hence
\begin{equation}
    \mathit{SNR}_R \leq \mathit{SNR}_A.
    \label{eq:snra}
\end{equation}
It is important to note, that in this receiver model no automatic gain control is performed, i.e. the output noise level remains constant for any input signal level below $S_\mathrm{max}$.

For modelling purposes, an ideal transmitter is assumed with nonidealities of the real transmitter lumped into the receiver model. For channel sounders, this is appropriate as transmitter and receiver are always considered together.


\section{Effects of Nonidealities}
For the derivation of bounds on the dynamic range of digital correlative channel sounders, (\ref{eq:snra}) is an important observation. It naturally follows that $\mathit{SNR}_A$ determines the achievable dynamic range $\mathit{DR}_A$. Furthermore, the achievable $\mathit{SNR}$ for a given receiver configuration cannot be exceeded by insertion of any additional elements into the signal chain, e.g. an LNA at the receiver input. It is a system parameter, determined by the signal bandwidth, intrinsic receiver noise and the received signal level at the edge of the linear region $S_{q,\mathrm{max}}$. As an example, $\mathit{SNR}_R$ in a (hypothetical) noiseless receiver with $S_{q,\mathrm{max}}=\SI{0}{\dbm}$ and $B=\SI{2}{\giga\hertz}$ measurement bandwidth cannot exceed $\mathit{SNR}_A=\SI{81}{\decibel}$. For frequency agile receivers, $\mathit{SNR}_A$ additionally is a function of the carrier frequency. Although the insertion of additional gain comes with a noise penalty, it may be beneficial nonetheless as will be discussed later.

\begin{figure}[tbp]
\centerline{\includegraphics[width=3.375in]{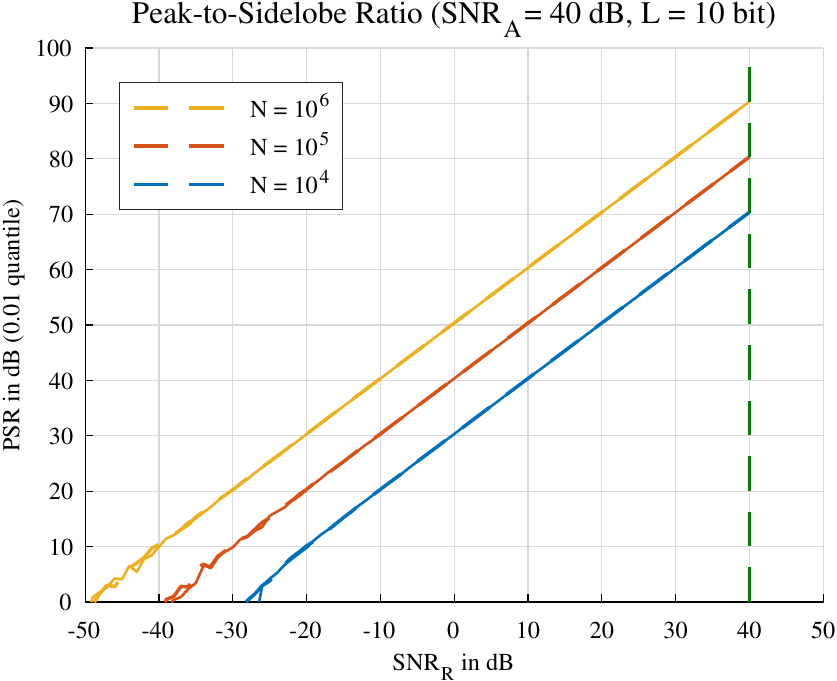}}
\caption{Simulated 0.01-quantile PSR of different length FZC sequences $s^N_{1}$ over received SNR for \SI{10}{\bit} quantization (solid lines) and achievable SNR \SI{40}{\decibel}. The results without quantization are plotted as dashed lines for comparison.}
\label{fig:psr_sim_length}
\end{figure}

\begin{figure}[tbp]
\centerline{\includegraphics[width=3.375in]{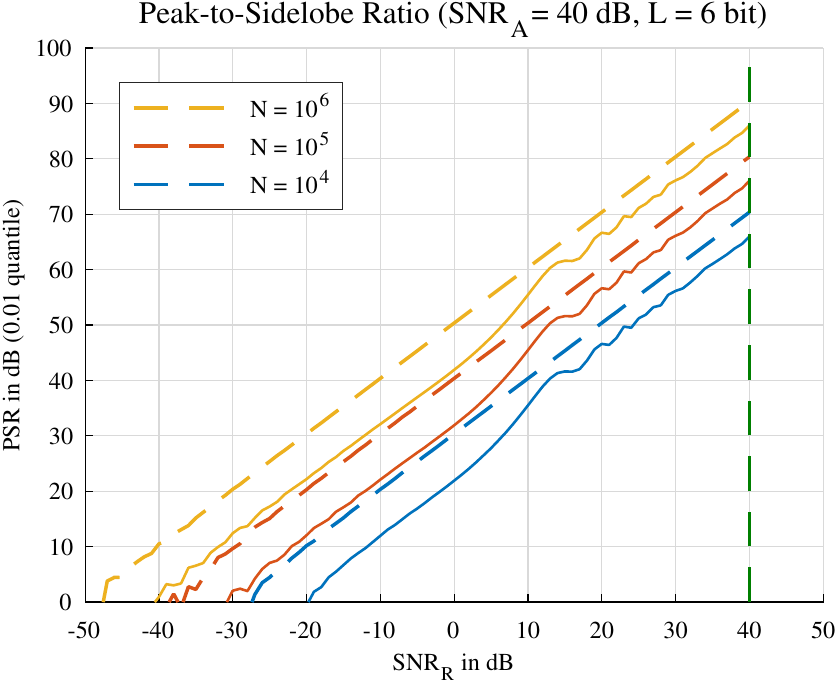}}
\caption{Simulated 0.01-quantile PSR of different length FZC sequences $s^N_{1}$ over received SNR for \SI{6}{\bit} quantization (dashed lines are without quantization effects)}
\label{fig:psr_sim_length_2}
\end{figure}
With $\mathit{SNR}_A$ given, it is possible to determine the achievable dynamic range for a particular FZC sounding sequence by means of $\mathit{PSR}$ simulation. Fig.~\ref{fig:psr_sim_length} shows the simulated $\mathit{PSR}$ over $\mathit{SNR}_R$ for three different choices of sequence lengths $N$ using a receiver with \SI{10}{\bit} quantization and $\mathit{SNR}_A=\SI{40}{\decibel}$. $\mathit{PSR}$ shows a linear dependency on $\mathit{SNR}_R$. For the given resolution of \SI{10}{\bit}, there is no significant difference between the quantized (solid lines) and continuous amplitude signals (dashed lines). In that case, the system performance is purely noise limited. The same simulation is shown for comparison in Fig.~\ref{fig:psr_sim_length_2} for \SI{6}{\bit} resolution. In this case, it can be noticed, that  peak-to-sidelobe ratio in the quantized case (solid lines) is decreased compared to the non-quantized case (dashed lines).

\begin{figure}[htbp]
\centerline{\includegraphics[width=3.375in]{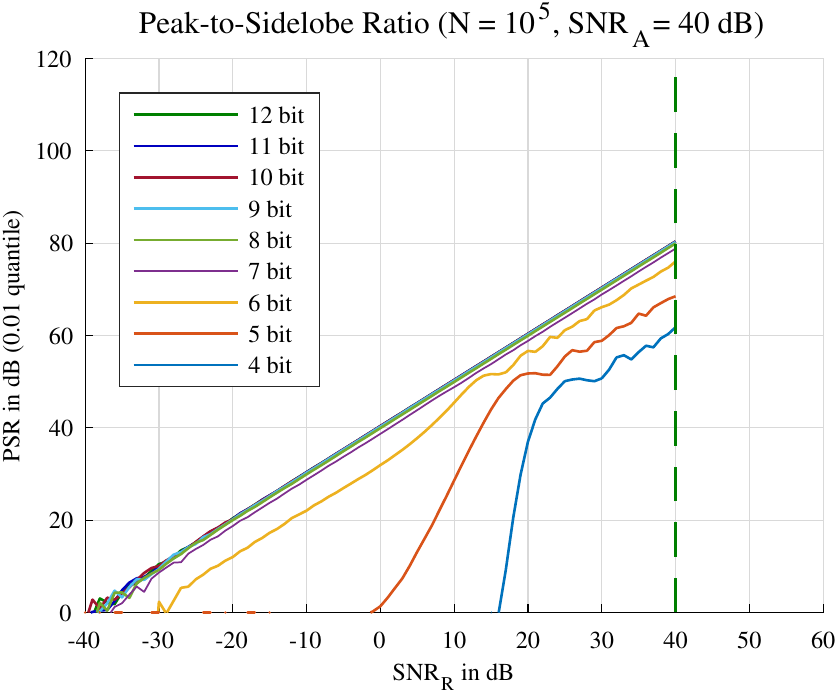}}
\caption{Simulated 0.01-quantile PSR of FZC sequence $s^{10^5}_{1}$ over received SNR for different receiver quantizer resolution and achievable SNR \SI{40}{\decibel}.}
\label{fig:psr_sim_quant_snra40}
\end{figure}

For further clarification, Fig.~\ref{fig:psr_sim_quant_snra40} shows the $\mathit{PSR}$ simulation using the same $\mathit{SNR}_A$, but for fixed sequence length $N=\num{1e5}$ and different quantizer resolutions $L$. Here, $\mathit{PSR}$ shows a threshold behavior w.r.t. $L$ below which the nonlinear effects of quantization become dominant. Above this threshold $L_\mathrm{min}$, which in this case is \SI{7}{\bit}, no significant gain is achieved by increased quantizer resolution.

\begin{figure}[htbp]
\centerline{\includegraphics[width=3.375in]{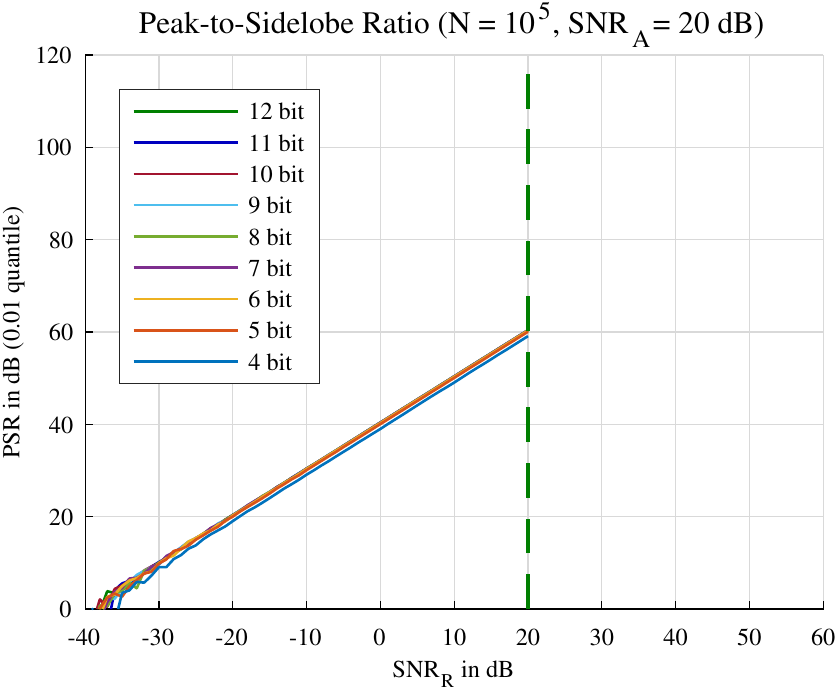}}
\caption{Simulated 0.01-quantile PSR of FZC sequence $s^{10^5}_{1}$ over received SNR for different receiver quantizer resolution and achievable SNR \SI{20}{\decibel}.}
\label{fig:psr_sim_quant_snra20}
\end{figure}
\begin{figure}[htbp]
\centerline{\includegraphics[width=3.375in]{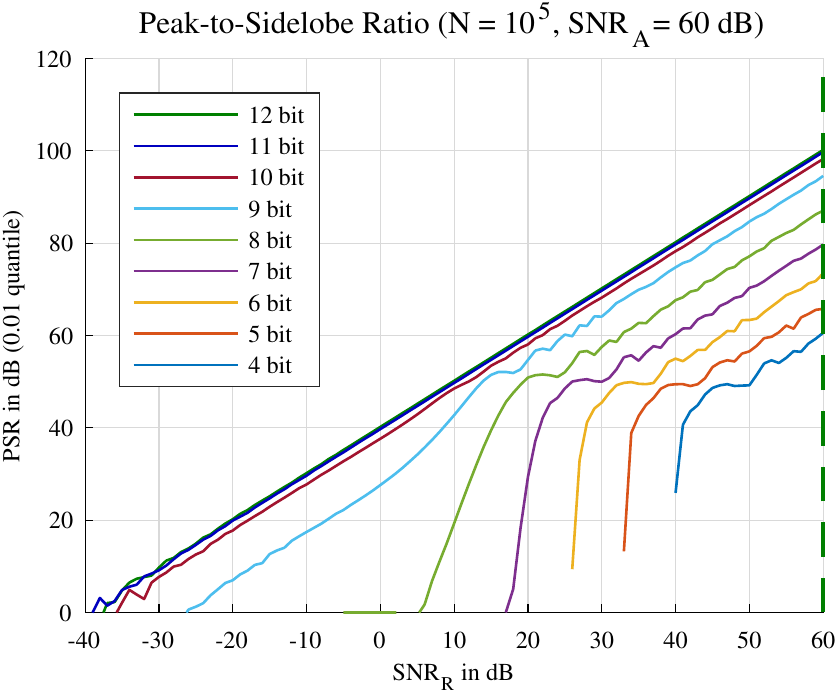}}
\caption{Simulated 0.01-quantile PSR of FZC sequence $s^{10^5}_{1}$ over received SNR for different receiver quantizer resolution and achievable SNR \SI{60}{\decibel}.}
\label{fig:psr_sim_quant_snra60}
\end{figure}

$L_\mathrm{min}$ is a function of $\mathit{SNR}_A$, which is illustrated by Figs.~\ref{fig:psr_sim_quant_snra20} and \ref{fig:psr_sim_quant_snra60}. In general, higher quantizer resolution is only beneficial in receivers with higher $\mathit{SNR}_A$. From the simulations, we may derive a general relationship between $\mathit{PSR}$, $G_\mathrm{proc}$ and $\mathit{SNR}_R$ and hence the dynamic range for FZC based digital correlative channel sounders. For sufficient quantization and typical $\mathit{SNR}_A$, $\mathit{PSR}_\mathrm{max}$ and therefore the maximum instantaneous dynamic range is given by:
\begin{equation}
    \mathit{IDR} \leq \mathit{PSR}_\mathrm{max} = \mathit{SNR}_R + G_\mathrm{proc} - \SI{9.7}{\decibel}.
    \label{eq:psr}
\end{equation}
The achievable dynamic range is given by
\begin{equation}
    \mathit{DR}_A = \mathit{SNR}_A + G_\mathrm{proc} - \SI{9.7}{\decibel}
\end{equation}

To conclude our discussion, we consider the absolute bounds on measurable path loss. The minimum measurable path loss is determined by maximum receiver input level $S_\mathrm{max}$ and transmit power $P_\mathrm{tx}$ with
\begin{equation}
    \mathit{PL}_{\mathrm{min},0} = P_{tx,\mathrm{max}} - S_\mathrm{max}.
\end{equation}

In order to maximize $\mathit{DR}_A$ at the minimal measurement distance $d_\mathrm{min}$ in a given scenario one may either
\begin{enumerate}
    \item increase the sounding signal's time-bandwidth product
    \item increase $P_\mathrm{tx}$ while $S < S_\mathrm{max}$ at $d_\mathrm{min}$
    \item if $\mathit{SNR}_R < \mathit{SNR}_{A,G}$ at $d_\mathrm{min}$ where $\mathit{SNR}_{A,G}$ is the achievable SNR of the receiver with an additional gain block at the input, use that gain block.
\end{enumerate}


\section{Practical Evaluation}
\begin{figure}[htbp]
\centerline{\includegraphics[width=3.375in]{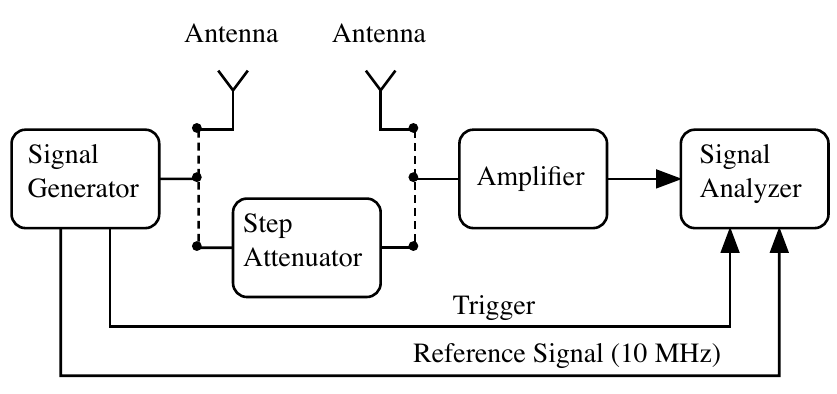}}
\caption{Measurement setup for the verification measurements.}
\label{fig:setup}
\end{figure}

To verify the results of the preceding discussion, two experiments were conducted in the measurement setup according to Fig.~\ref{fig:setup}.

In this setup, the signal generator is a R\&S\textsuperscript{\textregistered}SMW200A, the receiving signal analyzer is a R\&S\textsuperscript{\textregistered}FSW43, the step attenuator is a HP 84904K/84906K and the amplifier is a Millitech Ka-band LNA. Signal generator and signal analyzer share a \SI{10}{\mega\hertz} reference and the signal analyzer is triggered on the start of the transmit waveform. The transmit signals are FZC sequences $s^N_1$.

In the first experiment, a conducted measurement using the step attenuator to emulate a single-tap channel was performed without the additional LNA. In a first step, the noise power at receiver output $N_q$ (see Fig.~\ref{fig:frontend}) at \SI{25}{\giga\hertz} carrier frequency and \SI{2}{\giga\hertz} signal bandwidth was measured with the receiver input terminated in \SI{50}{\ohm}. Then, the transmit power was increased just before overload indication by the instrument and the input power $S_\mathrm{max}$ was measured by means of a R\&S\textsuperscript{\textregistered}NRP40SN power meter. With this information the achievable $\mathit{SNR_A}$ could be determined. After a back-to-back calibration to compensate for the system frequency response, different length FZC sequences were transmitted at step attenuator settings from \num{0} to \SI{90}{\decibel} and the received signals stored for post processing. In post processing, the CIRs were computed for each attenuator setting using a \SI{100}{\decibel}-sidelobe suppression Chebyshev window for pulse forming.

\begin{figure}[bp]
\centerline{\includegraphics[width=3.375in]{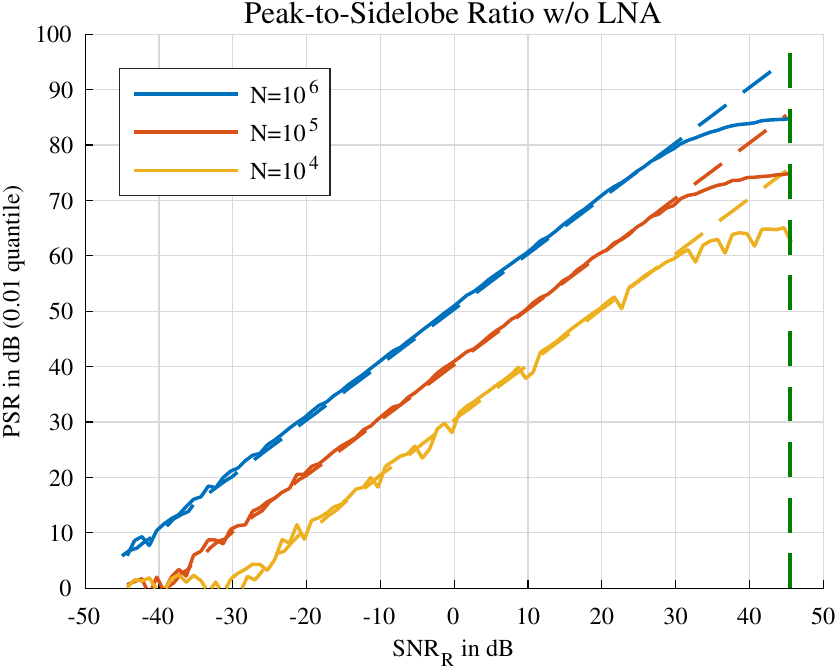}}
\caption{0.01-quantile PSR of different length FZC sequences $s^N_{1}$ over received signal-to-noise-ratio ($\mathit{SNR}_R$) in a conducted measurement at \SI{25}{\giga\hertz} with $\mathit{SNR}_A = \SI{45}{\decibel}$ (solid lines). $\mathit{SNR}_R$ was directly derived from the step attenuator settings with respect to $\mathit{SNR}_A$. Dashed lines show the corresponding simulation for comparison.}
\label{fig:psr_meas_length}
\end{figure}

\begin{figure}[tbp]
\centerline{\includegraphics[width=3.375in]{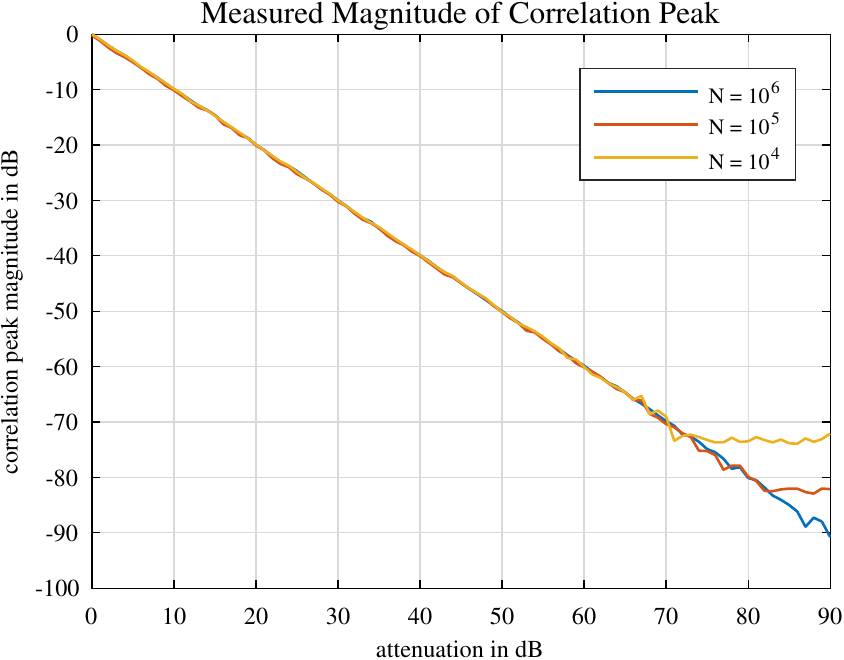}}
\caption{Measured correlation peak magnitude of different length FZC sequences $s^N_{1}$ at \SI{25}{\giga\hertz} over step attenuator settings.}
\label{fig:att_nolna}
\end{figure}

Fig.~\ref{fig:psr_meas_length} shows the measured PSR over $\mathit{SNR}_R$ for three different sequence lengths. An $\mathit{SNR}_A=\SI{45}{\decibel}$ for this receiver configuration was determined from the measured noise and signal levels. Dashed lines indicate the simulation results corresponding to this $\mathit{SNR}_A$ for comparison. The measurements show good agreement with the simulations, except for $\mathit{SNR}$ values close to $\mathit{SNR}_A$ where the behavior becomes nonlinear. This may be explained by frontend nonlinearities becoming stronger close to fullscale, meaning, that in real setups, there might be some saturation effect on the maximum available peak-to-sidelobe ratio $\mathit{PSR_A}$ solely depending on sequence length $N$ or processing gain $G_\mathrm{proc}$, respectively. On the other side, the measurement results indicate, that even with saturation, extraordinarily high $\mathit{PSR}$ values are practically achieved which could hardly be met by other measurement principles. Furthermore, it has to be noted that the degraded $\mathit{PSR}$ is merely a result of increased correlation errors in the side lobes and not that of amplitude compression as is shown by Fig.~\ref{fig:att_nolna}.

In Fig.~\ref{fig:att_nolna}, the peak magnitude of the measured CIR is plotted over the step attenuator settings, showing a purely linear behavior up to the noise floor.

Specifically, the \SI{10}{\decibel} spacing of lines for each sequence length in Fig.~\ref{fig:psr_meas_length} indicate that the theoretical correlation gain was fully realized. It also follows, that the effect of phase noise in this measurement configuration is negligible for the purpose of dynamic range evaluation.

\begin{figure}[htbp]
\centerline{\includegraphics[width=3.375in]{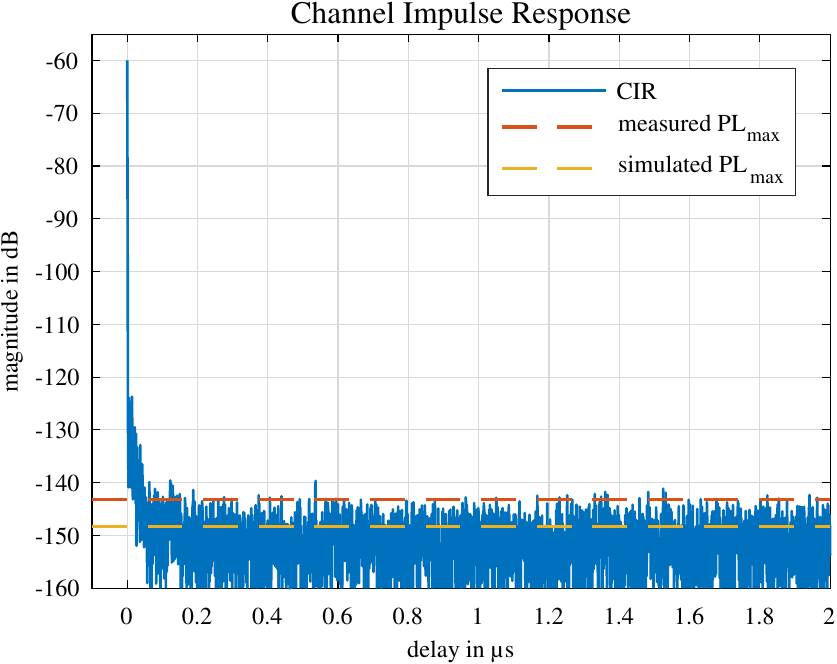}}
\caption{Measured CIR in a conducted setup with dashed lines indicating the measured and simulated PSR, respectively.}
\label{fig:cir_conducted}
\end{figure}

In the second experiment, the conducted setup was replaced by two custom-made omni-directional antennas at transmitter and receiver spaced \SI{1}{\meter} apart. In this experiment, the additional LNA was used. The carrier frequency was \SI{28}{\giga\hertz} and the transmitted signal a \SI{2}{\giga\hertz} bandwidth FZC sequence with $N=\num{2e5}$. Additionally, $K=10$ periods were recorded. Again, a conducted signal level and back-to-back calibration were performed, this time using a fixed \SI{60}{\decibel} attenuator to account for the expected \SI{61}{\decibel} free space path loss at \SI{1}{\meter} distance. After calibration, two measurements were taken, one for verification with the conducted connection including the attenuator and one over-the-air measurement. Post processing for CIR computation was identical with the first experiment.

\begin{figure}[bp]
\centerline{\includegraphics[width=3.375in]{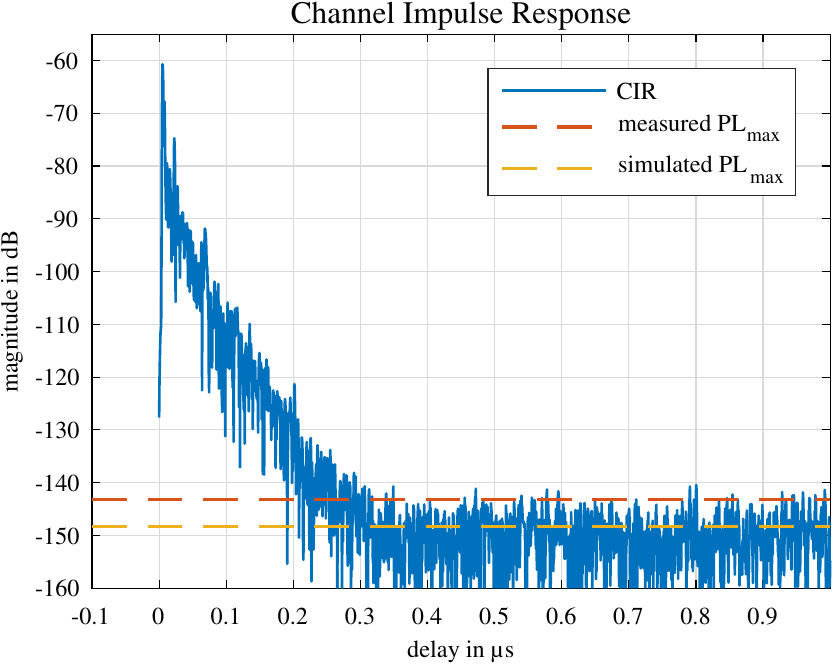}}
\caption{Measured CIR in an over-the-air setup with dashed lines indicating the measured and simulated PSR, respectively.}
\label{fig:cir_fspl}
\end{figure}

From the signal and noise level measurements, $\mathit{SNR}_A=\SI{34}{\decibel}$ was determined for this setup. Figs.~\ref{fig:cir_conducted} and \ref{fig:cir_fspl} show the measured CIRs for the conducted verification measurement and the over-the-air measurement, respectively. In the conducted case, the peak magnitude is \SI{-59.98}{\decibel} and in the over-the-air case \SI{-60.7}{\decibel}. In the over-the-air measurement, the line of sight path appears with a delay of \SI{5}{\nano\second} caused by \SI{20}{\centi\meter} antenna feed lines on each side which could not be included in the calibration. The $\mathrm{PSR}$ determined from the measured CIRs is \SI{83.2}{\decibel}. Since the receiver was calibrated at maximum input level, the minimum measurable path loss for this configuration is \SI{60}{\decibel}, resulting in a maximum measurable path loss of $\SI{60}{\decibel}+\SI{83.2}{\decibel}=\SI{143.2}{\decibel}$. The measured $\mathrm{PSR}$ is below the bound of \SI{88.3}{\decibel} given by (\ref{eq:psr}) which may be explained by the observed compression effect of $\mathrm{PSR}$ as illustrated in Fig.~\ref{fig:psr_meas_length}. The measurement results thus show good agreement with the simulations.


\section{Conclusion}
Establishing comparable, unambiguous quantitative measures for channel sounder performance is crucial for both system design and the interpretation of reported results. Addressing the quantification of path loss measurement capabilities, this paper gives general definitions for dynamic range, minimum and maximum measurable path loss as figures of merit that are valid irrespective of implementation.

For the class of digital correlative time domain channel sounders using Frank-Zadoff-Chu pulse compression waveforms for channel excitation, an upper bound on dynamic range is derived from a generalized instrumentation model. This facilitates verification of measurement data and allows for the derivation of a clear set of rules for the optimal parametrization of the channel sounder for a given measurement scenario.

The theoretical results have been verified using an instrument based millimeter-wave channel sounder in both conducted and over-the-air measurements.

\bibliographystyle{IEEEtran}
\bibliography{IEEEabrv,references}

\end{document}